\documentclass{article}
\usepackage{epsfig}

\begin{document}

{\large \bf Computer Simulation Center in Internet}
\begin{flushleft}
G.A. Tarnavsky\\
Institute of Computational Mathematics\\
and Mathematical Geophysics, Siberian Branch\\ of the Russian Academy of Sciences\\
Lavrentyev Avenue 6\\
630090, Novosibirsk, Russia\\
Email: Gennady.Tarnavsky@gmail.com\\

\bigskip
E.V. Vorozhtsov\\
Khristianovich Institute of Theoretical\\
and Applied Mechanics, Siberian Branch\\
of the Russian Academy of Sciences,\\
Institutskaya street 4/1\\
630090, Novosibirsk, Russia\\
Email: vorozh@itam.nsc.ru
\end{flushleft}


{\bf Abstract:} The general description of infrastructure and
content of SciShop.ru computer simulation center is given. This
resource is a new form of knowledge generation and remote
education using modern Cloud Computing technologies.

\medskip


\section{Introduction}

Recent achievements in the area of information technologies and
Internet give the reasons to believe that the new methods of the
organization of the scientific knowledge exchange process  have
already been formed by now, and in the near future, one may expect
a general passage from the conventional techniques of the
scientific knowledge dissemination via the paper journals to their
electronic counterparts.

The present paper is devoted to one of the advanced aspects of the
scientific knowledge dissemination -- a new form of the transfer
of the developed intellectual product, the program complexes for
solving various scientific and applied problems from the
developers to the users.

The process of the transfer of computational complexes
conventionally consisted  of the fact that the consumer (the
future user) acquired the codes (in some cases also the program
texts) from the code developers and then installed them on his
computer. To ensure a reliable and declared work of the program
complex the user must have a similar operation medium, including
the systems for visualizing the digital data.

The stage of the installation of computer programs is, on the
whole, very complex even when a support by the developer is
provided, and it often causes many difficulties the surmounting of
which requires considerable, intrinsically non-productive expenses
of intellectual efforts and time.

The new form of using the computational complexes frees from all
problems related to the installation of the acquired program
product.

A special site -- the Computer Simulation Center -- is created in
the Internet. This Center hosts the program complexes with all
their attributes: the preprocessor system for preparing the tasks,
the processor system for executing the tasks (executing the
computational operations), and the postprocessor system for the
output of the obtained information in the digital and graphical
forms.

The user must only formulate his own computational task by
performing the input of numerical data and starting its numerical
solution. He will be provided with the solution of his task upon
the termination of computations. Special services of the client's
support must provide a comfort of the visitor's stay at the
Center. In the economic terms, such a method of using the program
complexes means their leasing from the resource developers.

The Computer Simulation Center SciShop.ru is a pioneer of this new
direction of the development of the advanced information
technologies. The Center was created as a result of the execution
of the series of works (see, for example,
\cite{Tar09,Tar10,Tar05a,Tar05b,Tar07,TA08,AT07,Tar09b,EPE,Tar10b,ZhibT09}).

A general description and the infrastructure of the Center are
presented in \cite{Tar09,Tar10,Tar05a}, and its content, the
information-computational complexes for solving the tasks from a
number of scientific areas are presented in
\cite{Tar05b,Tar07,TA08,AT07,Tar09b}. At present, the Computer
Simulation  Center is functioning successfully in the Worldwide
web and is subject to its continuous modernization and development
(see \cite{Tar10b,ZhibT09}).

\begin{figure}[t]
 \epsfig{file=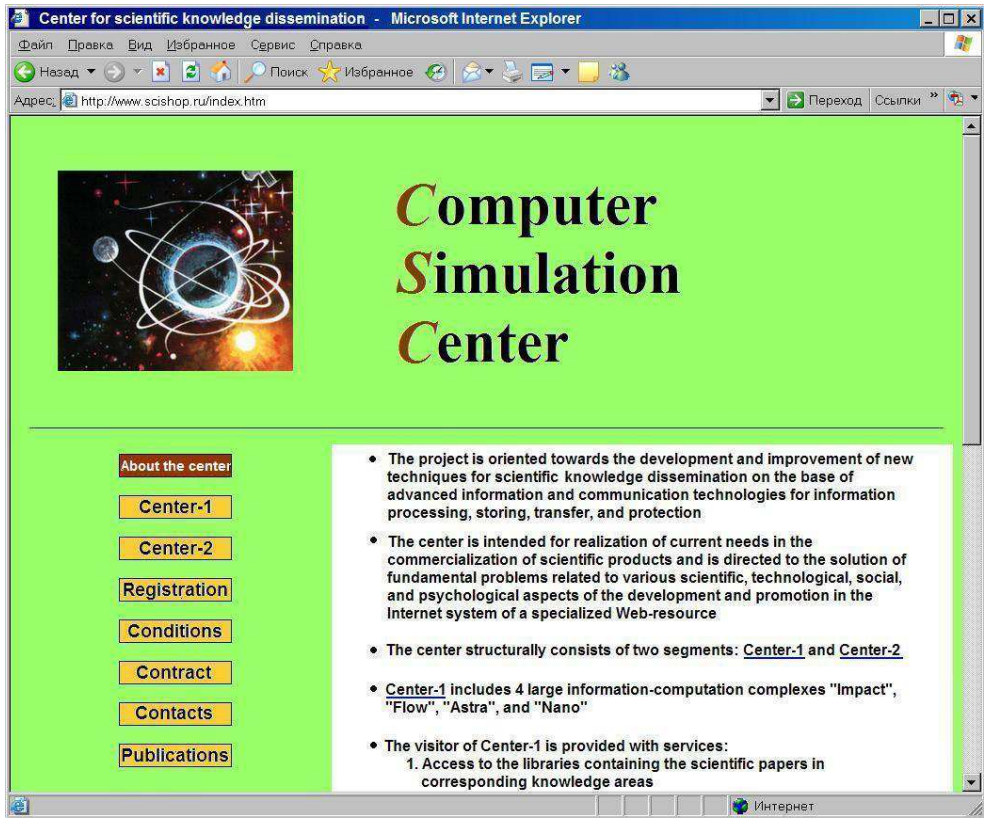, width=13cm}
 \caption{Main page of the Computer Simulation
 Center in Internet (a fragment).}\label{figtv1}
\end{figure}

This project is directed to the application of breakthrough
technologies in the domain of the development and improvement of
new techniques for scientific knowledge dissemination with the use
of the Worldwide web and is intended for a remote teaching of the
specialists, postgraduates, and students to the methods of
mathematical modelling and practical training in the solution of
scientific and applied problems. The teaching and training are
carried out on the basis of efficiently functioning program
complexes whose composition is continuously enriched.

\section{Computer Simulation Center SciShop.ru}
The project on the development of the Computer Simulation Center
(http://www. SciShop.ru) is oriented towards the development and
improvement of new techniques for scientific knowledge
dissemination on the basis of the advanced info-communication
technologies for the information processing, storing, transfer,
and protection.

\textbullet\,\, The Center is intended for the realization of
current needs in the commercialization of scientific products and
is directed to the solution of fundamental problems related to
various scientific, technological, social and psychological
aspects of the development and promotion in the system of the
specialized Internet Web resource.

\textbullet\,\, The Center structurally consists of two segments:
Center-1 and Center-2.

\textbullet\,\, Center-1 includes four large
information-computational complexes ``Shock'', ``Flow'',
``Astra'', and ``Nano''.

\textbullet\,\, The following services are provided for the
visitor of Center-1:

1. The access to the libraries containing the scientific articles
in corresponding knowledge areas.

2. The access to tabular and/or graphical databases containing the
results of computer simulation of the corresponding computer
tasks.

3. The access to processor systems enabling for the client himself
the organization and execution of computer simulation of the
problems, which are of interest to him.

4. The access to the locks to pass to the sites of the leading
Russian and foreign journals in corresponding knowledge areas.

5. The access to the segment ``Forum'' for obtaining the
consultations and discussion of problems.

\textbullet\,\, In the non-commercial regime, the visitor has the
access to the demo versions of the systems of Center-1. For a
full-scale access, one should register and perform the payment.

\textbullet\,\, The system of the Center  for accepting the
payments accepts the payment from any electronic payment systems
(WebMoney, Yandex.Money, E-gold, etc., which enter the Robokassa
consortium). The system for accepting payment with the use of the
bank credit cards and SMS messages of the cellular communications
has been developed and is now under verification.

\textbullet\,\, Center-2 is intended for positioning the program
developments in various branches of knowledge without any
limitations for the themes. The content of this section may be
augmented by the resources of any specialists having the copyright
for these information resources.

\textbullet\,\, All the specialists in the field of computer
simulation in any knowledge areas, which have the program
developments and wish to promote them, also on the onerous basis,
are invited for a cooperation with Center-2. The specialists
wishing to place their scientific products at Center-2 should
familiarize themselves with the conditions of their placement, get
in touch with the administration of the Center, and to sign a
Contract.

\textbullet\,\, One can familiarize himself in detail with all
aspects of the Center functioning
 in the special section ``Publications'' by using the corresponding hyperlink on the site Main page
(Fig. \ref{figtv1}).
\section{General characteristic of the arrangement\\ of computations at the center
and the advantages of the direct computer simulation in Internet}

The comfort of the client's stay in the Internet center and a
convenient form of using its segments: the bibliographic section,
databases containing already obtained tabular and graphical
information, and especially the processor systems are the most
important attributes of any program complex. Right this is usually
most difficult for the user.

The systems for the pre-processor preparation of tasks (the input
of parameters and the start-up of processor systems) are arranged
in a clear, convenient, and simplest form, which eliminates an
ambiguous interpretation and difficulties for the specialists
having even a small experience.

The operations with the processor systems are carried out by the
user in the regime of remote access in the Worldwide web on its
local  portal -- directly at the computer simulation center rather
than on its own computer after the installation of the complex.
This gives for the visitor of the Center a possibility to carry
out the study of the computational complex, to organize the
solution of the problem of interest, and to obtain the results of
computer studies.

\begin{figure}
 \epsfig{file=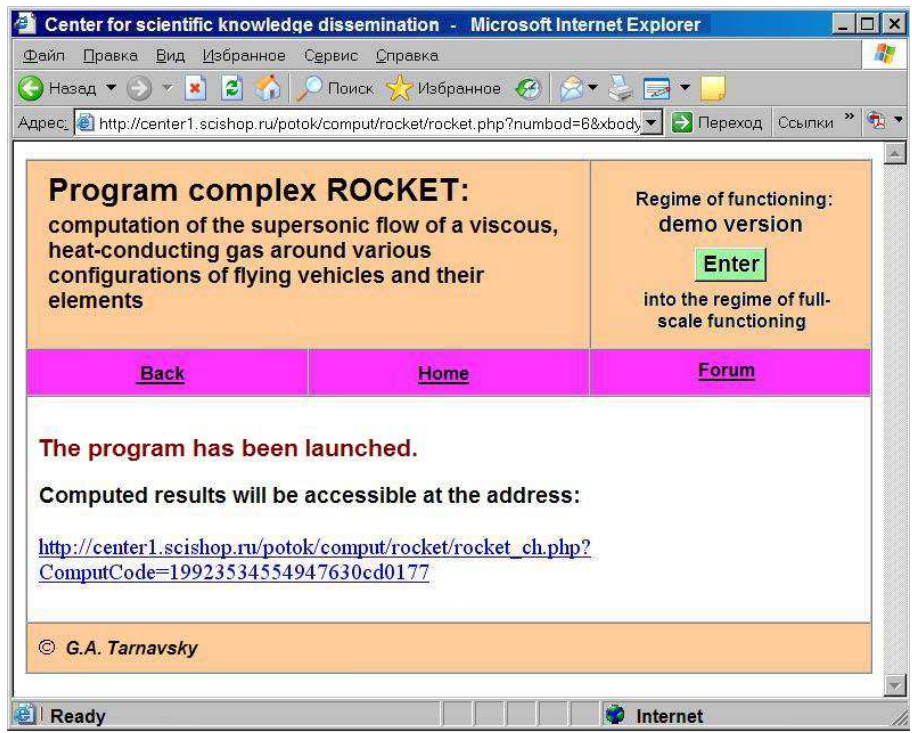, width=12cm}
\caption{Site page with the messages for the client that the
computation of the task formed by him has been
 launched, and its solution will be accessible at the indicated URL address.}\label{figtv2}
\end{figure}

The transfer of a computational complex usually consists of the
purchase of the license, documentation, and computer codes. After
that, the user performs the installation of the purchased product
on its own computer facility. As a rule, this occurs with large
difficulties, which may be due to various reasons, from using
different versions of the operating system to the peculiarities of
the supporting systems installed on computers of the seller and
the purchaser.

There are no such problems when the processor systems are placed
at the Center. All the interfaces have been debugged, well tested
and adjusted. The functioning of complexes is faultless in the
specified ranges of parameters variation (note that one can always
ask any question on the site forum and obtain the explanation).
The fact that the user is freed from the necessity of buying the
hardware (which is often very costly) necessary for performing the
needed computations is one more advantage of this approach.The
user in fact ``leases'' the hardware from the site developers only
for the time of the solution of his task.

We emphasize that such an efficient method of using the processor
complexes by the visitors of the Center has required the
development and implementation of original solutions. Since
neither of the Internet providers will permit the execution of
many and, possibly, long-term computations on his node, which
would take many resources and reduce the capacity of channels, one
would need the use of a different scheme for execution of
computations.

The client organizes at the Center his computational task (chooses
the processor system and inputs the parameters in it) and starts
up the computation on the system. The systems of this site segment
on the support of tasks pack the task into a file and send it via
the net to the computing center containing a number of computers,
including the multi-processor systems. The task is solved here,
the results (Fig. \ref{figtv2}) are forwarded either to the
Center, if the client waits for them, or to its home address in
the net. This scheme has shown itself very well during its beta
testing.

\section{The system for registering clients and commercial system of the center}
\begin{figure}[t]
 \epsfig{file=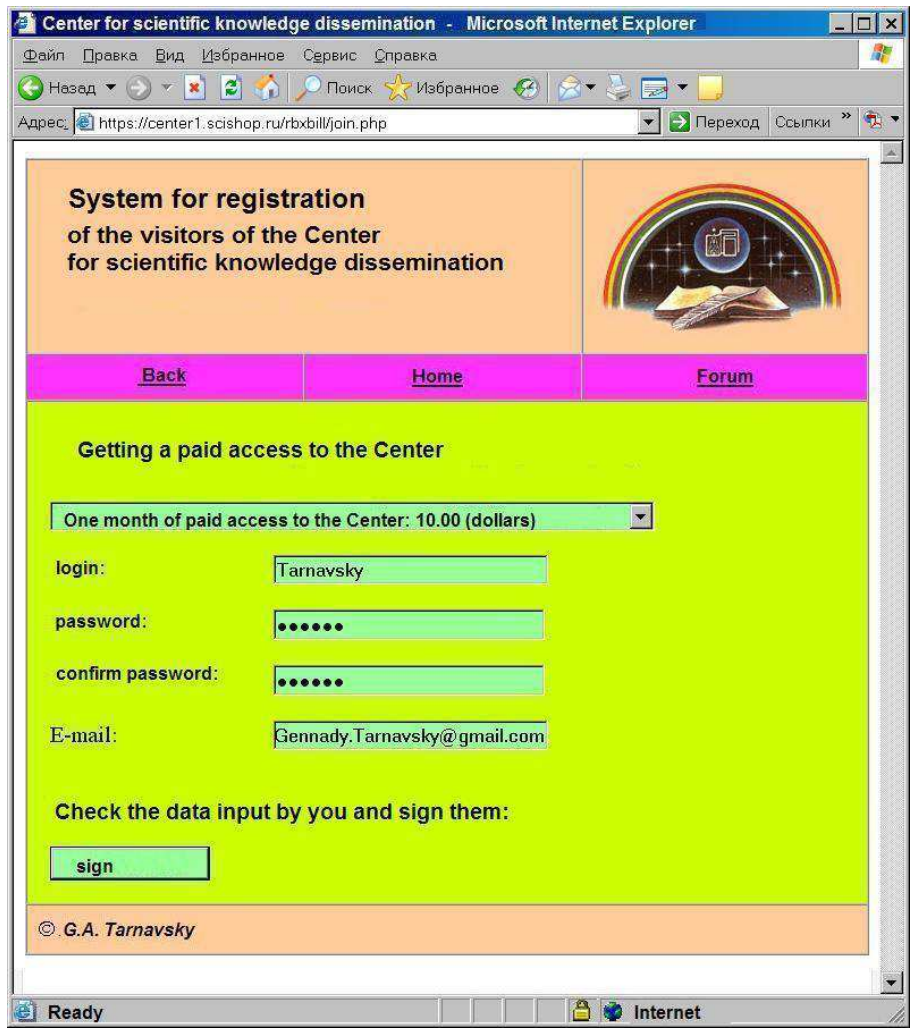, width=12cm}
\caption{Base page of the site section ``The system for
registration of the visitors of the Computer Simulation
Center''.}\label{figtv3}
\end{figure}

\begin{figure}[t]
 \epsfig{file=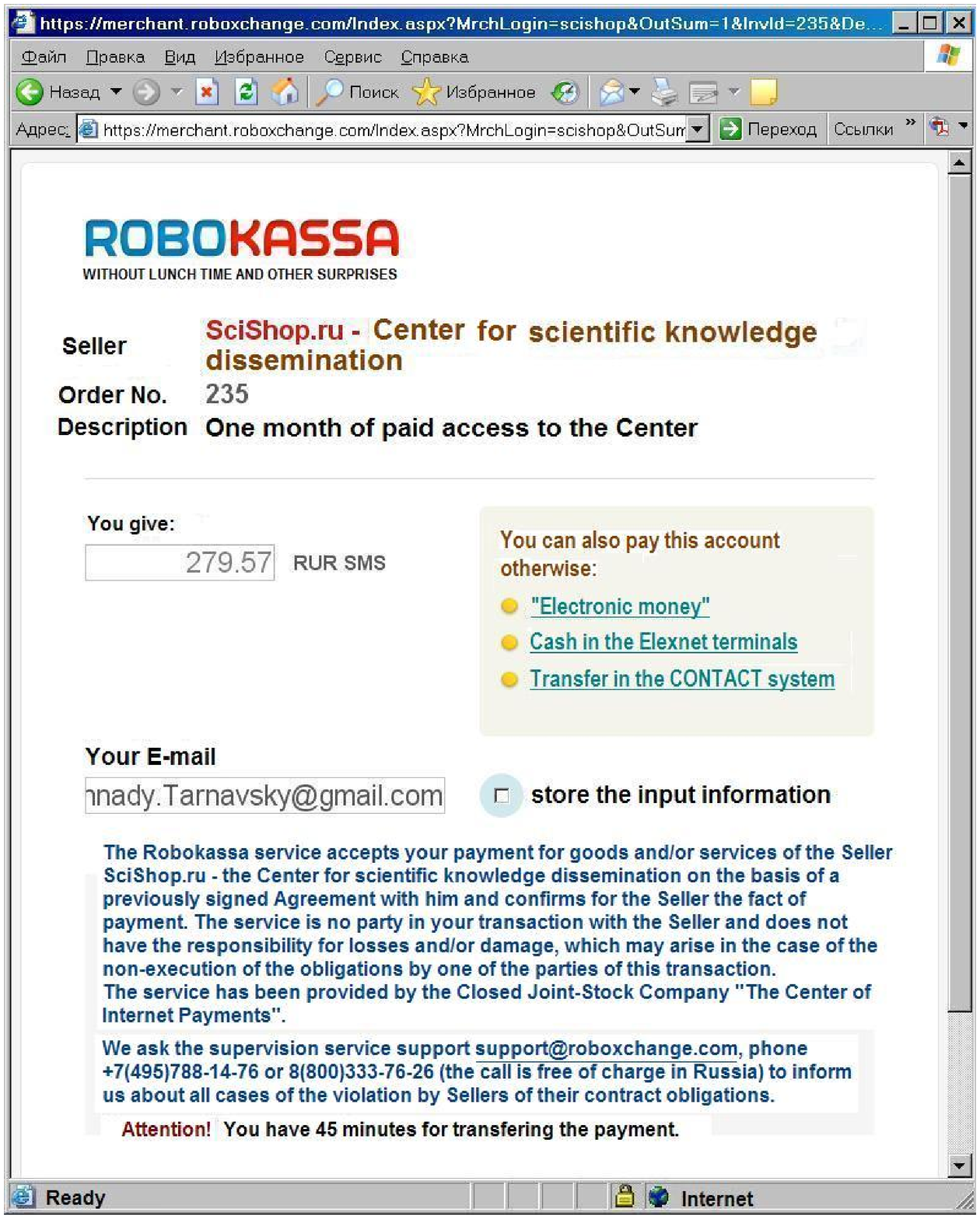, width=10cm}
\caption{One of the pages of the site section ``Transfer of user
charge via the SMS messages of cellular
communications''.}\label{figtv4}
\end{figure}

The program complexes of the Center may be used both free of
charge (the demo versions) and on a commercial basis. The access
to the regime of the full-scale functioning of the center is
realized after the visitor registration in the specialized ''book
of record'' (Fig. \ref{figtv3}) and after the visitor transfers
the user charge via the electronic payment systems.

The use of the specialized bank system ``Robokassa'' has been
organized (together with the development of necessary interfaces)
for the commercial segment of the center. This system enables the
use of more than 20 electronic payment systems (WebMoney,
Yandex.Money, Internet.Money, Internet.Groshi) as well as a number
of foreign electronic payment systems of the type E-Gold, PayPal,
MoneyBookers, EuroCash, etc. for the purpose of a significant
broadening of the scope of clients. A system has also been
implemented, which makes it possible to carry out the payments by
using the SMS cellular communications (Fig. \ref{figtv4}). The
safety of the passage of payments, the transparency of their
routing, the necessary messages for the client, the currency
conversion in different electronic payment systems are guaranteed
by special structures of the ``Robokassa'' system and have been
checked in the course of the beta testing  of the Computer
simulation center.

One should, however, emphasize that the profit earning is not set
at present as a predominant goal  of the present many-sided
project, which is a functional study of the problems of the
scientific knowledge dissemination based on advanced information
technologies.

\section{The center content}
The site SciShop.ru is a developing Center of computer simulation.
At present, it contains four working information-computation
complexes (ICCs):

- ``Shock'', the high-velocity internal aerodynamics: the
computation of shock-wave structures at the inlet to the diffuser
of a hypersonic scramjet engine;

- ``Flow'', the high-velocity external aerodynamics: the
computation of the flow around the objects in the atmospheres of
the Earth and Mars;

- ``Astra'', the computational astrophysics: simulation of the
dynamics of processes in the intergalactic gas and protoplanet
clouds;

- ``Nano'', microelectronics: computer support of the design of
nano-structured semiconductor materials.

Each of these complexes includes the bibliographic section,
tabular and graphical databases, which contain the results of the
computation of problems in their subject areas as well as the
processor systems, which enable the visitor to organize
independently the solution of the task of interest. All the
resources were created and improved in the course of the execution
of numerous computational experiments.
\begin{figure}[t]
 \epsfig{file=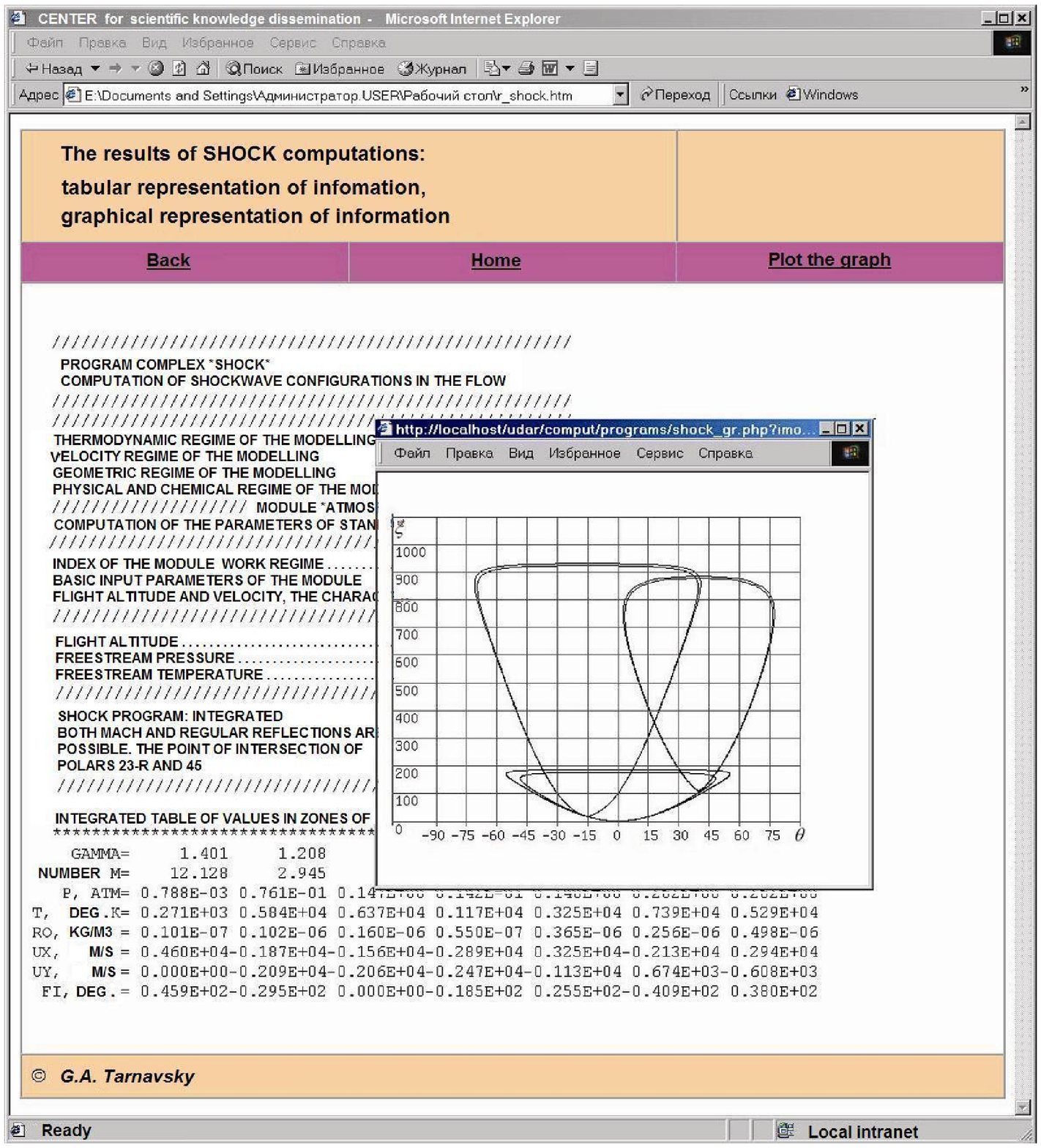, width=12cm}
\caption{Information-computation complex ``Shock''.
Tabular-digital and graphical representation of the
results.}\label{figtv5}
\end{figure}

ICC ``Shock'' (Fig. \ref{figtv5}). A wide spectrum of the
investigations of the shockwave flow patterns arising at the
hypersonic scramjet engine inlet was done in the works
\cite{Tar05a,Tar05b} with regard for the variation of the gaseous
medium properties on shocks. The configuration of shocks, the type
of their interaction (the Mach and/or regular type) and the flow
parameters between the shock fronts under the variation of the
flight altitude in the Earth atmosphere from 0 to 100 km, the
flight velocities from 1.5 to 7 km/s, the diffuser angles from 0
to 50 degrees were determined.

\begin{figure}[t]
 \epsfig{file=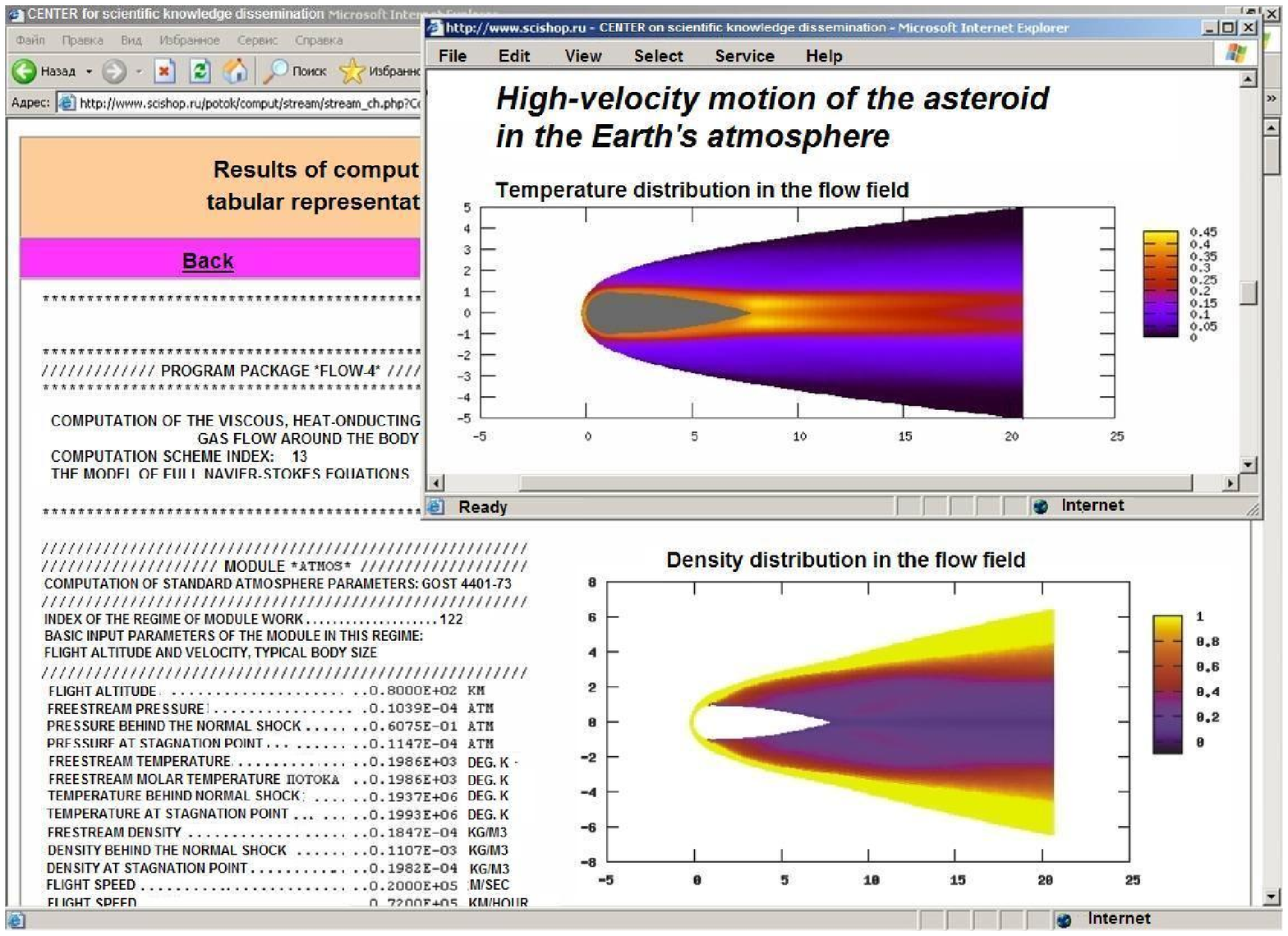, width=13cm}
\caption{Information-computation complex ``Flow''. Tabular-digital
and graphical representation of the results.}\label{figtv6}
\end{figure}

ICC ``Flow'' (Fig. \ref{figtv6}). The investigation of steady and
unsteady flows of both ideal gas (the model of Euler equations)
and viscous, heat-conducting gas (the model of full Navier--Stokes
equations) around the bodies of different configurations was
carried out in the works \cite{Tar07,TA08} on the basis of
specially developed methods and numerical algorithms within a wide
range of determining parameters. The flow structures near the
forebody, above the body lateral surface and the body base as well
as the flows in the near and far wake of the body were
investigated. The characteristic and peculiarities of these
structures were determined depending on flight regimes, including
the case of a localized energy supply to the free stream with the
formation of a complex flow pattern of the arising internal shock
waves, which are determined by the frequency of the sequence of
external source pulses, which may lead to resonance phenomena when
an insignificant power of pulses gives rise to a substantial
destabilization of the flow, high force and thermal loads on the
body surface.

The complex enables the modelling of the aerodynamics of the
supersonic and hypersonic flight in the Earth's atmosphere at the
altitudes from 0 to 100 km within the speed range from 0.8 to 10
km/s as well as, to a certain accuracy (in accordance with
available data), a high-velocity flight in the Mars' atmosphere.

In these works, the most important problems of the adequacy of the
mathematical model and the algorithms and codes realizing it to
the occurring physical process were studied at a qualitatively new
level. Three segments of this question, which are related to the
nonuniqueness, were considered: the analytic solutions of the
Euler equations, numerical solutions of the full Navier--Stokes
equations, and the symmetry loss in symmetric problems. The
processes of the evolution of the pattern of the flow around the
body with the possibility of a passage from one solution branch to
another  were studied, and the attraction basins of solutions were
determined. The trajectories of the computation motion from the
starting solution to the final (steady, quasi-steady, unsteady
aperiodic) solution in the space of solutions were studied.

\begin{figure}[t]
 \epsfig{file=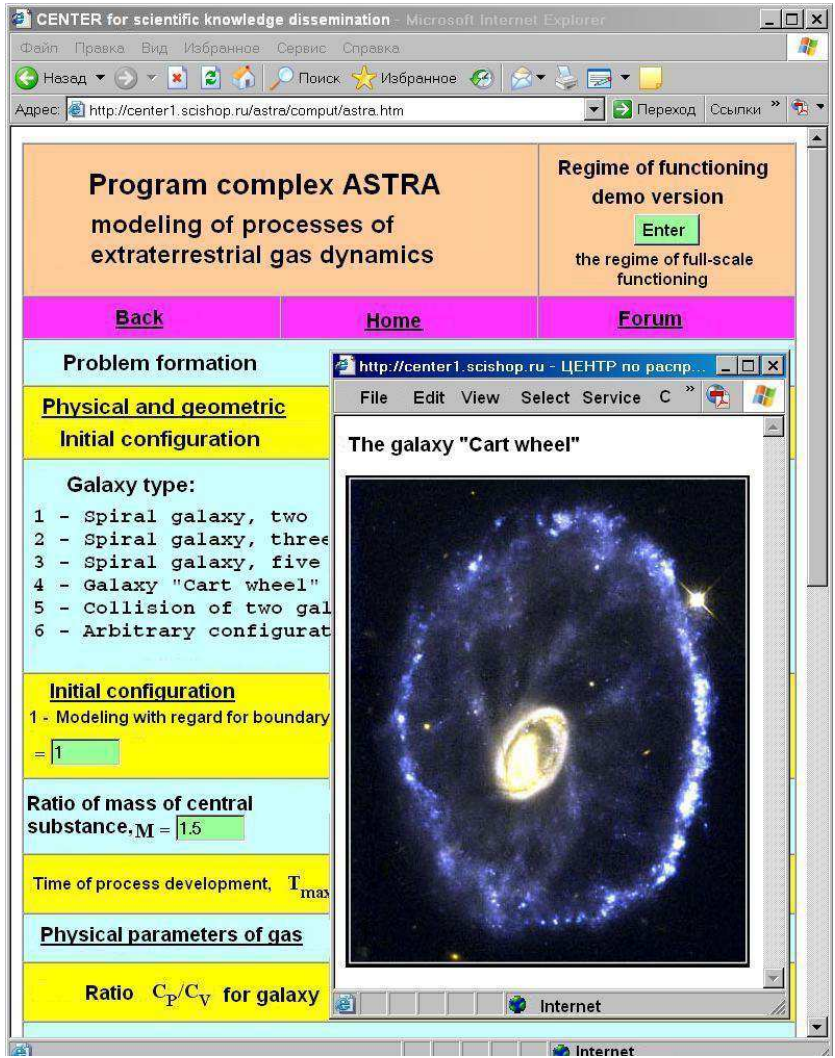, width=11cm}
\caption{Information-computation complex ``Astra''. Page for the
preparation of tasks and graphical representation of the
results.}\label{figtv7}

\vspace*{-5mm}
\end{figure}

ICC ``Astra'' (Fig. \ref{figtv7}). The physical, mathematical, and
computational problems of modelling the unsteady three-dimensional
problems of the extraterrestrial gas dynamics were considered and
analyzed in the work \cite{AT07}. The system of the Euler gas
dynamic equations, which was completed by the force and energy
components to model the deviation of the equation of state from
the ideal one, the heat-transfer processes (heat conduction,
convection, and radiation), gravitation (the gravity field of the
point mass and the self-gravitation of a distributed gaseous
cloud), was used as the governing system of the differential
equations of the mathematical model. The modelling was carried out
on the basis of the principle of the decomposition of the complete
problem into several sub-problems corresponding to different
physical processes. The structuring of the computational complex
into several autonomous segments, in its turn, corresponded to
this decomposition. This ensures the possibility for extension and
supplement of the package of computer programs. The series of the
computations of problems on the motion of shock waves and
expansion waves in intergalactic gaseous media, on the
gravitational collapse of motionless and rotating gaseous clouds,
the recession of a gaseous cluster, which model the formation and
explosion of proto-stars, were done. A thorough verification of
the theoretical method, the computational algorithm, and complex
of computer codes was done for a comprehensive analysis of their
properties (the accuracy of computation and the speed of the
computational process).

\begin{figure}[t]
 \epsfig{file=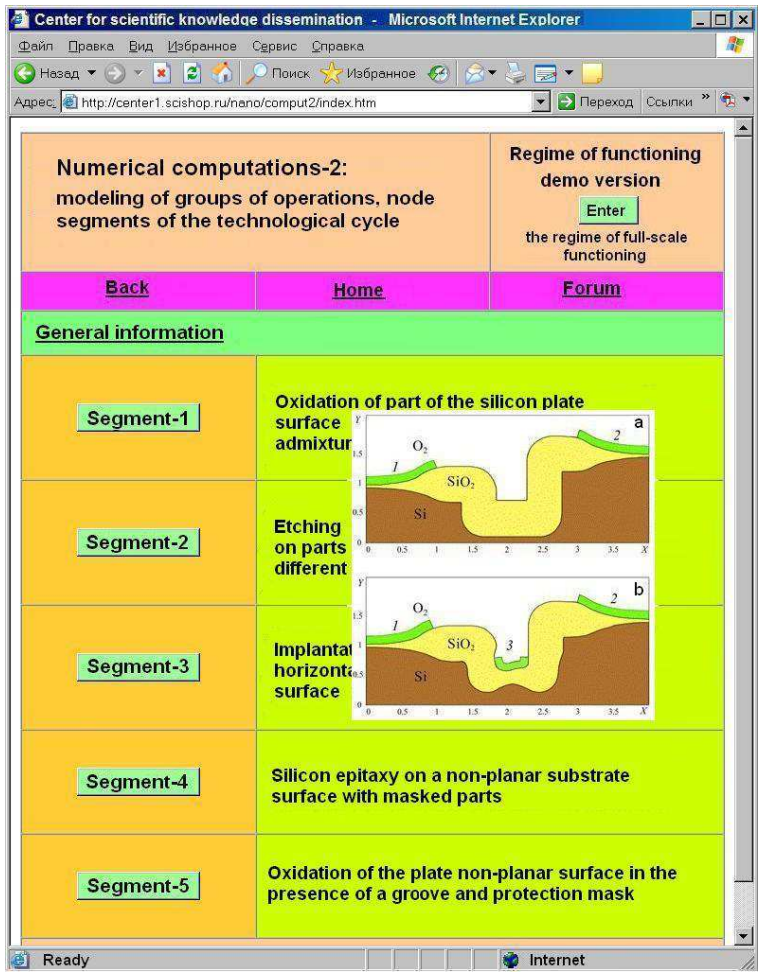, width=10cm}
\caption{Information-computation complex ``Nano''. The page of a
task for simulation of the groups of technological operations  and
graphical representation of the results.}\label{figtv8}

\end{figure}

ICC ``Nano'' (Fig. \ref{figtv8}). Theoretical methods were
developed in the work \cite{Tar09b} for mathematical modelling of
a number of physical-chemical and mechanical processes of the
technological cycle of the production of new semiconductor
materials, in particular, the motion of the oxidation wave in
silicon, including the case of the availability of technological
masks ensuring special configurations of the interfaces
``material/oxide'' with the formation of multiply connected
regions.  These methods were the basis for developing the
efficient numerical algorithms and the complex of computer
programs.

Special methods, high-accuracy algorithms, and computer codes were
developed for computing the physical processes of the segregation
of dopant donor and acceptor admixtures (boron, antimony, and
arsenic) at the oxidation wave front in a free-crystalline and
prestressed silicon \cite{EPE}.

The computer simulation of the formation of specific
nanostructures -- narrow localized zones of an elevated electric
conductivity of the  n- and p-types was conducted.

The complex enables the design of nano- and microelectromechanical
systems (diodes, capacitors, transistors, etc.) entering the
large, very large, and ultra large integral circuits.

A many-sided experience obtained in the course of the works was
implemented in the algorithms of program complexes, which are
granted to the visitors of the Center \cite{Tar10b,ZhibT09} for
solving their own tasks in the corresponding knowledge areas.

\section{Conclusion}

The infrastructure and content of the Computer simulation center
SciShop.ru, the pioneer of a new form of the scientific knowledge
dissemination, were briefly considered in the present paper. This
center is intended for a direct use of program complexes for the
mathematical simulation of processes in various scientific areas,
and it provides the possibility of a direct execution of
computations in Internet in the remote access regime. Such a form
has wide prospects of the application in scientific research and
applied developments as well as for a remote teaching of
specialists, postgraduates, and students.



\label{lastpage}
\end{document}